\documentclass[prb, aps, twocolumn, amsmath,amssymb,amsfonts]{revtex4}
\usepackage{graphicx}
\usepackage{graphics}
\usepackage{dcolumn}
\usepackage{bm}
\usepackage{amssymb}
\usepackage{amsmath}
\usepackage{amsfonts}
\usepackage{epsfig}

\newcommand {\bra} [1] {\langle #1 |}
\newcommand {\ket} [1] {| #1 \rangle}

\newcommand {\dbkt} [2] {\langle #1 | #2 \rangle}
\newcommand {\tbkt} [3] {\langle #1 | #2 | #3 \rangle}
\newcommand {\pd} [2] {\frac{\partial #1}{\partial #2}}
\newcommand {\td} [2] {\frac{d #1}{d #2}}

 \newcommand {\beq}{\begin{equation}}
\newcommand {\eeq}{\end{equation}}
\newcommand {\bea}{\begin{eqnarray}}
\newcommand {\eea}{\end{eqnarray}}

\begin{document}
\title{External gates and transport in biased bilayer graphene}

\author{Dimitrie Culcer}
\author{R.~Winkler}

\affiliation{Advanced Photon Source, Argonne National Laboratory,
Argonne, IL 60439.}

\affiliation{Northern Illinois University, De Kalb, IL 60115.}

\begin{abstract}
  We formulate a theory of transport in graphene bilayers in the
  weak momentum scattering regime in such a way as to take into
  account contributions to the electrical conductivity to leading
  and next-to-leading order in the scattering potential. The
  response of bilayers to an electric field cannot be regarded as a
  sum of terms due to individual layers. Rather, interlayer
  tunneling and coherence between positive- and negative-energy
  states give the main contributions to the conductivity. At low
  energies, the dominant effect of scattering on transport comes
  from scattering within each energy band, yet a simple picture
  encapsulating the role of collisions in a set of scattering times
  is not applicable. Coherence between positive- and negative-energy
  states gives, as in monolayers, a term in the conductivity which
  depends on the order of limits. The application of an external
  gate, which introduces a gap between positive- and negative-energy
  states, does not affect transport. Nevertheless the solution to
  the kinetic equation in the presence of such a gate is very
  revealing for transport in both bilayers and monolayers.
\end{abstract}
\date{\today}

\maketitle

\section{Introduction}

The manufacture of carbon monolayers in the laboratory has generated
unprecedented excitement. The unusual and sometimes baffling
properties of graphene have exposed the community to novel science
and supplied ideas for technological innovation. \cite{net07r} This
achievement was swiftly followed by the reliable manufacture of
graphene bilayers and multilayers, which have become independent
research areas in their own right. The electronic properties of
bilayer graphene \cite{McCannFalko} have been the subject of recent
reviews \cite{Netobil, Netobias} but the topic has been studied much
less than single graphene layers. Large mobilities recently observed
\cite{Hopeful} are among many factors justifying experimental and
theoretical interest in bilayers. Also notable in these systems is a
Berry phase of $2\pi$ accompanying an unconventional quantum Hall
effect \cite{2piBerry} and predictions of Andreev reflection
\cite{Andreev} and superfluidity \cite{Superfl} among other effects.
\cite{Choi, Semenov, Ivar, Mkhi, KoshDeloc, KatsZeroE, Nemec,
Netoferrobil1, Netoferrobil2} Studies of graphene bilayers have
focused on transport, \cite{AdamBoltz, Katsmin, KoshFP, Mogh, Nicol,
Cserti, Kech, Gorb, Snyman, Aber} compressibility, \cite{ViolaComp}
impurities, \cite{Bena, KatsImp, NilssImp} electron-electron
interactions, \cite{Nilssee, Violaee, Hwang, Chakra} and band and
electronic structure. \cite{Wang, HongkiBS, CastroBS, McCann, Li,
Henrik}

Graphene bilayers are interesting from a physics point of view
because they are not merely the sum of two layers. Tunneling between
layers is characterized by the large parameter $t_\perp$ which is
comparable to the Fermi energy at carrier densities $n$ well beyond
$10^{12}$~cm$^{-2}$. The energy spectrum of bilayers consists of
four bands, two with positive energy and two with negative energy.
The lower positive-energy band and the higher negative-energy band
touch at $k=0$. Although at low densities only one of these latter
two bands is occupied, depending on whether the sample is doped with
electrons or holes, a pseudospin cannot be defined in the same
manner as in single graphene layers. This is due to the presence of
the large tunneling parameter, which indicates a nontrivial
interplay of carriers from both layers, particularly important in
steady-state processes. Furthermore, a gate potential opens a gap in
the energy spectrum, \cite{CastroBS} and independent control of this
gap and of the carrier density can be achieved by using separate top
and back gates. \cite{Oostinga}

This article presents a thorough investigation of transport in
graphene bilayers, constructing a compact and straightforward
framework for analyzing the structure of the steady-state density
matrix in an electric field. The formalism takes the density
operator and quantum Liouville equation as its starting point,
treating all terms in the Hamiltonian on the same footing and using
realistic scattering potentials. Calculation of the electrical
current reveals the complex physics underlying bilayer transport.
The conductivity contains a term which is a function of the carrier
density $n$ and is inversely proportional to the impurity density
$n_i$, similar to the usual conductivity of metals and
semiconductors. This term also depends on $t_\perp$, which can be
thought of as indicating coherence between layers, while its
dependence on $n$ is different for long-range and short range
impurities and it is in general not $\propto n$. The concept of a
characteristic momentum scattering time is useful as an order of
magnitude, but it is not possible to assign scattering times to
carriers in different bands. Furthermore, an electric field couples
positive and negative energy states, resulting in a term in the
conductivity similar to that in single layers but of different
magnitude because of the different winding number associated with
energy dispersion in bilayers. It must also be borne in mind that
positive- and negative-energy states involve carriers from both
layers. In graphene monolayers the conductivity due to the coupling
of positive- and negative-energy states is renormalized by
scattering to twice its original value. In bilayers the effect of
scattering on this term is considerably smaller than in monolayers.
This term also depends on the order of limits, and work to date has
not been able to extract its definite value. Fortunately, we will
show that biased bilayer graphene ought to provide an answer. The
external gate makes a contribution to the conductivity through an
off-diagonal (Hall) term which would appear to exist without a
magnetic field. This term also depends on the order of limits, but
it is the sole contribution to the off-diagonal conductivity. Since
crystal symmetry and Onsager relations imply that an off-diagonal
conductivity cannot exist without a magnetic field this must
indicate the correct and unambiguous order of limits. If any doubts
persist experiment can surely resolve this issue, given that there
is only one potential contribution to the off-diagonal conductivity.
We note that Boltzmann transport theory has been formulated
thoroughly for unbiased bilayer graphene. \cite{AdamBoltz} In this
work we wish to consider both the ordinary (``Boltzmann'')
contribution to the conductivity $\propto n/n_i$ and the
contribution independent of $n$ and $n_i$ on the same footing. We
also aim to determine unambiguously the role of the gate and the
lessons to be learned from it.

We focus on extrinsic graphene in the weak momentum scattering
regime, $\varepsilon_F \gg \hbar/\tau$, where $\varepsilon_F$ is the
Fermi energy and $\tau \propto n_i^{-1}$ is a characteristic
momentum scattering time (for strong momentum scattering
$\varepsilon_F \ll \hbar/\tau$.) Extrinsic graphene refers to the
case $n>0$ and may be in either the weak or the strong momentum
scattering regime. Intrinsic graphene refers to the case $n=0$ where
the Fermi energy lies at the point where the bands touch, and is by
definition in the strong momentum scattering regime. Enormous
strides in sample quality make transport in the weak momentum
scattering regime a timely undertaking. \cite{net07r, Netobil,
Netobias} We assume low temperatures, where scattering due to
charged impurities is important and may dominate, and where
electron-electron scattering plays a secondary role. In the regime
of weak momentum scattering studied here quantum interference
effects are also expected to be negligible. \cite{Kechedzhi} We
stress once more that a conductivity independent of $n$ and $n_i$
was measured experimentally by taking $n\rightarrow 0$. The value
obtained is characteristic of the \emph{strong} momentum scattering
regime and is referred to as the \emph{minimum conductivity}.
\cite{Hopeful} At the same time, theoretical research on
\emph{clean} samples finds an additional conductivity independent of
$n$ and $n_i$. It is this latter term, rather than the minimum
conductivity, that is discussed in our work.

The outline of this article is as follows. In Section II we
construct the kinetic equation for bilayer graphene taking the
quantum Liouville equation as our starting point. We discuss in
detail the role of a general elastic impurity potential in this
kinetic equation specific to bilayers. In Section III we apply this
equation to the study of transport in unbiased bilayers and identify
contributions to the conductivity $\propto n/n_i$ and independent of
$n$ and $n_i$. In Section IV we discuss the role of the gate and
show that, if treated in a naive manner, it can appear to yield an
off-diagonal conductivity in the absence of a magnetic field. We
discuss the implications of this finding for transport in graphene
bilayers and monolayers.

\section{Kinetic equation}

The formalism parallels that used in Ref.~\onlinecite{UsGraphene}
and its exposition below is correspondingly abbreviated. The system
is described by a density operator $\hat \rho$. Evaluation of
$\hat\rho$ in the steady state allows one to calculate expectation
values such as that of the velocity operator. Very generally,
$\hat\rho$ obeys the quantum Liouville equation
\begin{equation}
\td{\hat\rho}{t} + \frac{i}{\hbar} \,
\bigl[\hat H + \hat{H}^E + \hat{U}, \hat \rho \bigr]
= 0.
\end{equation}
Here $\hat{H}$ is the band Hamiltonian, $\hat{H}^E =e{\bm
E}\cdot\hat{\bm r}$ the interaction with the external electric field
$\bm{E}$, and $\hat{U}$ the impurity potential. We project the
Liouville equation onto a set of time-independent states of definite
wave vector $\{ \ket{{\bm k}s} \}$, in which $\bra{{\bm k}s}
\hat\rho \ket{{\bm k}'s'} \equiv \rho^{ss'}_{{\bm k}{\bm k}'} \equiv
\rho_{{\bm k}{\bm k}'} $ and similarly $H_{{\bm k}{\bm k}'}$,
$H^E_{{\bm k}{\bm k}'}$, and $U_{{\bm k}{\bm k}'}$. For bilayers the
index $s$ runs from 1 to 4 as will be shown below. We refer to
$\rho_{{\bm k}{\bm k}'}$ as the density matrix. Matrix elements of
$H_{{\bm k}{\bm k}'} = H_{{\bm k}} \, \delta_{{\bm k}{\bm k}'}$ are
diagonal in ${\bm k}$ but off-diagonal in $ss'$, and similar for
$H^E_{{\bm k}} $. Matrix elements of $U_{{\bm k}{\bm k}'}$ are
off-diagonal in ${\bm k}$. Elastic scattering is assumed and the
average of terms $U_{{\bm k}{\bm k}'} U_{{\bm k}'{\bm k}}$ in the
disorder potential over impurity configurations is $n_i
|\bar{U}_{{\bm k}{\bm k}'}|^2 /V$, where $V$ is the crystal volume
and $\bar{U}_{{\bm k}{\bm k}'}$ the matrix element of the potential
of a single impurity. $\rho_{{\bm k}{\bm k}'}$ has a part $f_{\bm
k}$ diagonal in ${\bm k}$, and a part off-diagonal in ${\bm k}$. We
will be interested in $f_{\bm k}$ since most operators related with
steady-state processes are diagonal in ${\bm k}$. From the Liouville
equation an effective equation is derived for $f_{\bm k}$ in the
first Born approximation, valid for $\varepsilon_F \tau/\hbar \gg 1$
\begin{equation}
  \pd{f_{{\bm k}}}{t} + \frac{i}{\hbar} \, 
  [H_{{\bm k}}, f_{{\bm k}}] + \hat{J}(f_{\bm k}) = \Sigma_{\bm k},
\end{equation}
where the source term $\Sigma_{\bm k} = (e{\bm
E}/\hbar)\cdot(\partial f_0/\partial{\bm k})$, the equilibrium
density matrix $f_0(H_{\bm k})$ is given by the Fermi-Dirac function
and the scattering term takes the form
\begin{equation}
\hat{J} (f_{\bm k}) =
(n_i/\hbar^2)\, \int_0^\infty dt' \, [\hat{\bar U}, e^{- i \hat H
t'/\hbar}[\hat {\bar U}, \hat f]\, e^{i \hat H t'/\hbar}]_{{\bm
k}{\bm k}}.
\end{equation}

The low-energy bilayer graphene Hamiltonian is \cite{Netobil}
\begin{equation}
\label{eq:ham}
H_{{\bm k}} = \begin{pmatrix} 0 & \hbar v
ke^{i\phi} & t_\perp & 0 \\ \hbar v ke^{-i\phi} & 0 & 0 & 0 \\
t_\perp & 0 & 0 &  \hbar v ke^{-i\phi} \\ 0 & 0 &  \hbar v
ke^{i\phi} & 0
\end{pmatrix}.
\end{equation}
Along the diagonal are two $2\times 2$ submatrices which represent
the Hamiltonians of individual layers, in which $v \approx 1.1
\times 10^6$~m$^{-1}$ stands for the (constant) Fermi velocity of
graphene. These layers are coupled by the interlayer tunneling
parameter $t_\perp \approx 0.3$~eV. Although Eq.\ (\ref{eq:ham})
does not include the so-called \textit{trigonal warping} terms, this
model captures most of the important physics. \cite{Netobil} If we
define $\lambda_k = \sqrt{t_\perp^2 + 4\hbar^2 v^2 k^2}$, the energy
eigenvalues can be labeled as $\varepsilon_{k1} = \frac{1}{2}
(\lambda_k + t_\perp) = - \varepsilon_{k4}$ and $\varepsilon_{k2} =
\frac{1}{2} (\lambda_k - t_\perp) = - \varepsilon_{k3}$, independent
of the direction of ${\bm k}$. The energies $\varepsilon_{k1}$ and
$\varepsilon_{k2}$ are positive, whereas $\varepsilon_{k3}$ and
$\varepsilon_{k4}$ are negative. The two bands $\varepsilon_{k2}$
and $\varepsilon_{k3}$ touch at $k=0$. We work henceforth in the
basis of eigenstates of $H_{{\bm k}}$ with the eigenvectors labeled
$\ket{u_{{\bm k}s}}$ so that $\ket{{\bm k}s} = e^{i \bm{k} \cdot
\hat{\bm{r}}} \ket{u_{\bm{k}s}}$. In this basis the Hamiltonian is
diagonal and has the form $H_{{\bm k}} = {\rm
diag}(\varepsilon_{k1}, \varepsilon_{k2}, \varepsilon_{k3},
\varepsilon_{k4})$. Nevertheless, one must be careful in writing
down the kinetic equation in this basis. As the basis functions,
namely the eigenvectors of $H_{\bm k}$, depend on the magnitude and
direction of the wave vector ${\bm k}$, the ordinary derivative with
respect to ${\bm k}$ must be replaced by the covariant derivative.
The action of this covariant derivative for example on $f_0$ is
given by $Df_0/D{\bm k} = \partial f_0/\partial{\bm k} - i
[\bm{\mathcal R}, f_0]$. The connection matrix $\bm{\mathcal R}$
which enters the covariant derivative has elements
$\mathcal{R}_{ss'} = \dbkt{u_{{\bm k}s}}{i \partial u_{{\bm
k}s'}/\partial{\bm k}}$. It is easiest to work out the derivatives
with respect to the magnitude $k$ and polar angle $\theta$ of the
wave vector,
\begin{subequations}
\begin{eqnarray}
\mathcal{R}^k  & = & \frac{i\hbar v t_\perp}{\lambda_k^2}
\begin{pmatrix} 0 & 0 & - 1 & 0 \\ 0 & 0 & 0 & 1 \\
 1 & 0 & 0 & 0 \\
0 & - 1& 0 & 0
\end{pmatrix} \\ [1ex]
\mathcal{R}^\theta  & = & \frac{1}{k} \begin{pmatrix}
1 & -\frac{\hbar vk}{\lambda_k} & 0 & 0 \\
-\frac{\hbar vk}{\lambda_k}  & 1 & -1 &  0 \\
0 & -1 & 1 &  -\frac{\hbar vk}{\lambda_k} \\
0 & 0 &  -\frac{\hbar vk}{\lambda_k} & 1
\end{pmatrix} 
\end{eqnarray}
\end{subequations}
These expressions will be needed when constructing the velocity
operator, as well as when determining the source term in the kinetic
equation in an electric field. Lastly, since the Hamiltonian of
graphene bilayers and single layers does not depend on the spin of
the particles, final results will contain a factor of 2 from the sum
over the spin, as well as an additional factor of 2 which takes into
account the twofold valley degeneracy of graphene. Therefore the
final expressions for the conductivity must be multiplied by an
overall factor of 4.

The matrix elements of the scattering potential $\hat{\bar U}$ due
to a single impurity in the basis of $H_{\bm k}$ eigenstates are
\begin{equation}
\tbkt{{\bm k}s}{\hat {\bar U}}{{\bm k}'s'} = \mathcal{U}_{{\bm k}{\bm k}'} \,
M_{{\bm k}{\bm k}'}^{ss'} \equiv
\mathcal{U}_{{\bm k}{\bm k}'} \, M_{{\bm k}{\bm k}'},
\end{equation}
where $\mathcal{U}_{{\bm k}{\bm k}'}$ are the matrix elements of
$\hat{\bar U}$ between plane wave states. The band indices $s$ and
$s'$ will be henceforth suppressed and quantities such as $M_{{\bm
k}{\bm k}'}$ will be treated as matrices in the subspace spanned by
the four bands under consideration (with the band index $s$ the same
as that introduced above.) The scattering term $\hat{J} (f_{\bm k})$
appearing in the kinetic equation simplifies considerably if we
assume that the tunneling parameter $t_\perp \gg \hbar v k$. This
assumption is valid for carrier densities up to approximately
$10^{12}$~cm$^{-2}$. At these densities only one of the bands is
occupied: for electron (hole) doping this is the band labeled 2 (3).
We expand $M_{{\bm k}{\bm k}'}$ in the ratio $\hbar vk/t_\perp$ up
to order 1, and we find that the term of order 1 vanishes identically.
We label the
incident wave vector by ${\bm k}$, the outgoing wave vector by ${\bm
k}'$ and the polar angle of the outgoing wave vector ${\bm k}'$ by
$\theta'$. If we define $\gamma = \theta' - \theta$ as the relative
angle between incident and outgoing wave vectors, $M_{{\bm k}{\bm
k}'}$ has the simple diagonal form
\begin{equation}
M_{{\bm k}{\bm k}'} = e^{-i\gamma} \, {\rm diag} (1, \cos \gamma,
\cos \gamma, 1).
\end{equation}

\section{Transport without gates}

The most transparent solution to the kinetic equation is found by
dividing all matrices in the problem into a diagonal part, denoted
by the superscript $d$, and an off-diagonal part, denoted by the
superscript $od$. In the case of the density matrix the diagonal
part $f^d_{\bm k}$ represents the fraction of carriers which are in
eigenstates of $H_{\bm k}$, while $f^{od}_{\bm k}$ is the fraction
of carriers which are a continually changing mixture of eigenstates
of $H_{\bm k}$. Diagonal matrices commute with $H_{{\bm k}}$ while
off-diagonal matrices do not. The kinetic equation is
correspondingly divided into equations for the diagonal and
off-diagonal parts of the density matrix, which are coupled by
scattering
\begin{subequations} \label{dod}
\begin{eqnarray}
\label{eq:d}
\pd{f^d_{\bm k}}{t} + \hat{P}^d \hat{J} (f_{\bm k}) &
= & \Sigma^d_{\bm k} \\ [1ex]
\label{eq:od}
\pd{f^{od}_{\bm k}}{t} + \frac{i}{\hbar}\, [H_{\bm k}, f^{od}_{\bm k}] &
= & \Sigma^{od}_{\bm k} - \hat{P}^{od} \hat{J} (f_{\bm k}).
\end{eqnarray}
\end{subequations}
$\hat{P}^d$ and $\hat{P}^{od}$ are projection operators which single
out the diagonal and off-diagonal parts of matrices respectively. To
solve these equations, we search for terms in the density matrix of
lowest orders in $\hbar/(\varepsilon_F\tau)$ or equivalently those
terms of lowest orders in $n_i$. Inspection of Eq.\ (\ref{eq:d})
shows that, due to the absence of the commutator $[H_{\bm k},
f^d_{\bm k}] $, $f^d_{\bm k}$ starts at order $n_i^{-1}$ while the
leading term in $f^{od}_{\bm k}$ is independent of $n_i$ (in other
words order zero). \cite{UsGraphene} For weak momentum scattering
therefore we only need to consider the effect of the scattering term
acting on the diagonal part $f^d_{\bm k}$ of the density matrix.
This reduces to the simple form
\begin{equation}
\hat{P}^d \hat{J} (f^d_{\bm k}) = \frac{n_i \lambda_k}{8 \hbar^3 v^2}
\int \frac{d\theta'}{2\pi} \, |\mathcal{U}_{{\bm k}{\bm k}'}|^2 F^d (\gamma)
\,(f^d_{\bm k} - f^d_{{\bm k}'}),
\end{equation}
with the diagonal matrix $F^d (\gamma)$ given by 
\begin{equation}
F^d (\gamma) = \mathrm{diag} (2, 1 + \cos2\gamma, 1 + \cos2\gamma, 2). 
\end{equation}
This matrix is due solely to the overlap of eigenstates at different
wave vectors. The equation for the diagonal part $f^d_{\bm k}$ of
the density matrix in the steady state, in which the time derivative
can be dropped, reduces to
\begin{equation}
\frac{n_i \lambda_k}{8\hbar^3 v^2}
\int \frac{d\theta'}{2\pi} \, |\mathcal{U}_{{\bm k}{\bm k}'}|^2 F^d (\gamma) \,
(f^d_{\bm k} - f^d_{{\bm k}'}) = \Sigma^d_{\bm k}.
\end{equation}
The solution to this equation is found simply as $f^d_{\bm k} =
\Sigma^d_{\bm k}\tau_m$, where $\tau_m$, which plays the role of a
momentum scattering time, is a diagonal matrix given by
\begin{equation}
\tau_m^{-1} = \frac{n_i \lambda_k}{8\hbar^3 v^2} \int \frac{d\theta'}{2\pi} \,
|\mathcal{U}_{{\bm k}{\bm k}'}|^2 F^d(\gamma) (1 - \cos\gamma)
\end{equation}
with the matrix elements $\tau_{m11} = \tau_{m44} = \tau_+$ and
$\tau_{m22} = \tau_{m33} = \tau_-$. This allows to write the
electric-field-induced correction to the diagonal part of the
density matrix as
\begin{equation}
f^d_{\bm k} = - \frac{2 e\hbar v^2{\bm E}\cdot{\bm k}}{\lambda_k} \,
\mathrm{diag} (\tau_+ \, \delta_1, \tau_- \delta_2, - \tau_- \, \delta_3,
- \tau_+ \, \delta_4), 
\end{equation}
where we have used the abbreviation $\delta_s \equiv
\delta(\varepsilon_{ks} - \varepsilon_F)$. We would like to
stress that, although $\tau_m$ has the units of time, its elements
do not correspond to actual scattering times and their
energy-dependence does not come explicitly through any of the band
energies, but rather through $\lambda_k$, which represents the
difference between band energies. The form of $\lambda_k$ implies
that the dominant contribution to $\tau_m$ comes from $t_\perp$.
Therefore, whereas $f^d_{\bm k} \propto n_i^{-1}$ does indicate that
scattering tends to keep the Fermi surface near equilibrium, an
expression for this function in terms of a momentum relaxation time
cannot be formulated in bilayers.

The next step is to evaluate the off-diagonal part $f^{od}_{\bm k}$.
Firstly, the effective source term in Eq.\ (\ref{eq:od}) contains
the off-diagonal projection of the scattering term acting on
$f^d_{\bm k}$, given by $\hat{P}^{od} \hat{J} (f^d_{\bm k})$. We
find that this projection produces corrections of order $(\hbar v
k/t_\perp)^2$ and therefore can be omitted. Secondly, when carrying
out the time integral that determines the correction due to
$\Sigma^{od}_{\bm k}$ we allow the field to have a time dependence
$e^{-i \omega t}$, taking the limit $\omega \rightarrow 0$ at the
end. (An unphysical negative conductivity may be obtained if this
procedure is not followed. \cite{UsGraphene}) $f^{od}_{\bm k}$ is
found using the time evolution operator $e^{-iHt'/\hbar}$
\begin{equation}
f^{od}_{\bm k} = \lim_{\eta \rightarrow 0} \int_0^\infty \!\!\! dt'
\, e^{-\eta t'}
e^{i \omega (t' - t)}  e^{-iH_{\bm k}t'/\hbar} \Sigma^{od}_{\bm k}
\, e^{iH_{\bm k}t'/\hbar},
\end{equation}
with $\eta > 0$ a regularization factor. The time integral results
in a series of $\delta$-functions of the form
$\delta(\varepsilon_{ks} - \varepsilon_{ks'})$, with $s
\ne s'$. The only $\delta$-function that can take nonzero values is
$\delta(\varepsilon_{k2} - \varepsilon_{k3})$, due to
the two bands that touch at $k = 0$. The problem of finding
$f^{od}_{\bm k}$ therefore reduces to finding its matrix elements in
the subspace spanned by $\ket{u_{{\bm k}2}}$ and $\ket{u_{{\bm
k}3}}$. The only nonzero matrix element is
\begin{equation}
f^{od}_{{\bm k}23} = - \frac{i \pi e{\bm E} \cdot \hat{\bm \theta} } {2k} \,
\lim_{\omega \rightarrow 0} [f_0 (\varepsilon_{k2})
- f_0 (\varepsilon_{k2} - \hbar \omega)] \,
\delta \bigg(\varepsilon_{k2} - \frac{\hbar \omega}{2} \bigg)
\end{equation}
Thus the bands that touch at $k=0$ give $f^{od}_{\bm k}$, much like
in single-layer graphene. Finally, the diagonal projection
$\hat{P}^d \hat{J}(f^{od}_{\bm k})$, which would act as an effective
source in the equation for $f^d_{\bm k}$, is also of order $(\hbar v
k/t_\perp)^2$ and is omitted.

We determine separately the contributions to the electrical
conductivity due to each term in the density matrix for electron
doping, for which only the band labeled 2 is occupied. We require
the expectation value of the current operator $\hat{\bm j} =
-e\hat{\bm v}$, where $\hat{\bm v} = (1/\hbar) DH_{{\bm k}}/D{\bm
k}$ is the velocity operator in the basis of eigenstates of $H_{{\bm
k}}$, and the covariant derivative $D/D{\bm k}$ has been defined
above. $f^d_{\bm k}$, the fraction of carriers in eigenstates of
$H_{{\bm k}}$, yields (per valley and spin)
\begin{equation} 
\sigma^{xx}_d = \frac{e^2 \zeta}{h} \, \frac{n\pi \hbar^2 v^2 t^2_\perp}
 {(4\pi^2 \hbar^4 v^4)\, n^2 + (4 \pi \hbar^2 v^2 t^2_\perp)\, n + t_\perp^4}
\end{equation}
with the dimensionless quantity $\zeta = \tau_- \lambda_k/\hbar$. In
bilayers the screening wave vector is independent of the number
density, \cite{Hwang} and $\zeta \propto n$ for long-range
scatterers and is a constant for short-range scatterers. For
short-range impurities, at low densities $\sigma^{xx}_d \propto n$
(Ref.\ \onlinecite{AdamBoltz}) as in single-layer graphene;
\cite{net07r} but as the density increases nonlinear terms in $n$
become more pronounced. For long-range impurities $\sigma^{xx}_d
\propto n^2$ at low densities. The dependence of $\sigma^{xx}_d$ on
$t_\perp$ indicates that $\sigma^{xx}_d$ is due to carriers from
both layers. Nevertheless, the leading term $\propto t_\perp^{-2}$,
implying that at low densities interlayer tunneling hinders the
transport of charge.

For a nonzero chemical potential $\mu$, $f^{od}_{\bm k}$ gives (per
valley and spin)
\begin{equation}
\sigma^{xx}_{od} = \frac{\pi e^2}{4h} \lim_{\omega \rightarrow 0}
\bigg[ \frac{1}
{1 + e^{-\beta(\mu/2 + \hbar \omega/4) }}
- \frac{1} {1 + e^{-\beta(\mu/2 - \hbar \omega/4) }} \bigg],
\end{equation}
where $\beta = 1 / kT$. To obtain the dc result at $T=0$, one must
take the limits $T \rightarrow 0$ and $\omega \rightarrow 0$; yet
the result depends on the order in which these limits are taken. If
$T \rightarrow 0$ first the result is $\pi e^2/(4h)$, whereas if
$\omega \rightarrow 0$ first the result is zero. The same conundrum
is present in single layers of graphene. \cite{UsGraphene} At
present it is not clear whether this term is finite, and the theory
presented in this paper up to now does not offer an indisputable
solution. So far neither theory nor experiment can disambiguate this
issue. Experiment could provide a conclusive answer if clean samples
with zero carrier density were available, but that remains a
daunting task. For weak momentum scattering, $\sigma^{xx}_{od}$ is
considerably smaller than $\sigma^{xx}_d$ and cannot be extrapolated
conclusively from a plot of $n/n_i$. Fortunately however, the
analysis presented in the following section on biased bilayer
graphene shows that, if we consider the conductivity due to a gate
the answer can be found.

\section{Gate effect on transport}

We have determined so far the steady-state density matrix in
unbiased graphene bilayers. We study next the interesting case of
biased bilayer systems, in which the gap can be modified by the
application of an external gate voltage $V_g$. The gate voltage gives rise
to an additional term $H^g_{\bm k}$ in the Hamiltonian, which in the
basis of eigenstates of $H_{{\bm k}}$ has the form
\begin{equation}
H^g_{\bm k} = -\frac{eV_g}{2\lambda_k} \begin{pmatrix} 0
  & - \hbar v k  & 0 & t_\perp \\ - \hbar v k
  & 0 & - t_\perp& 0 \\ 0 & - t_\perp & 0 & - \hbar v k  \\
  t_\perp & 0 & - \hbar v k  & 0
\end{pmatrix}
\end{equation}
We treat the gate potential $-eV_g$ in first-order perturbation
theory. It is easily checked that to first order in $|eV_g|$ and $\hbar
v k/t_\perp$ the contribution of $H^g_{\bm k}$ to the scattering
term $\hat{J}(f_{\bm k})$ in the kinetic equation is zero. The gate
gives rise to an additional source term in the kinetic equation,
which takes the form
\begin{equation}
\Sigma^g_{\bm k} = \frac{e{\bm E}}{2\hbar}\, \cdot \frac{D}{D{\bm k}}\,
\bigg\{ H^g_{\bm k},  \bigg(\pd{f_0}{k}\bigg)\bigg(\pd{H}{k}\bigg)^{-1} \bigg\}
- \frac{i}{\hbar} \, [H^g_{\bm k}, f_{\bm k}] .
\end{equation}
The bracket $\{\cdot\}$ denotes the symmetrized dot product $\{ {\bm
a}\cdot{\bm b} \} \equiv {\bm a}\cdot{\bm b} + {\bm b}\cdot{\bm a}$.
Once again, the covariant derivative $D/D{\bm k}$ appears instead of
the ordinary derivative. We proceed exactly as before in order to
find the additional correction to the density matrix due to the
gate, which we call $f^g_{\bm k}$. The nonzero diagonal terms in
$f^g_{\bm k}$ are
\begin{widetext}
\begin{equation}
  f^g_{{\bm k}22} = - f^g_{{\bm k}33}
  = -\frac{\pi e^2 V_g t_\perp \tau_- {\bm E} \cdot \hat{\bm \theta} }
    {4\hbar k \lambda_k}
    \lim_{\omega \rightarrow 0}
    {\textstyle
    \left[  f_0 \left(\frac{ \hbar \omega}{2}\right)
      - f_0 \left(\frac{-\hbar \omega}{2}\right) \right]
     \delta\left(\varepsilon_{k2} - \frac{\hbar \omega}{2}\right)}.
\end{equation}
\end{widetext}
Several other terms either vanish as $\omega \rightarrow 0$ or give
zero contributions to steady-state expectation values after
integration over wave vector. These terms are omitted here for
simplicity and without loss of generality.

Following the same procedure as in the unbiased case discussed
above, we find that the nonequilibrium correction $f^g_{\bm k}$ to
the density matrix due to the gate gives only an
\textit{off-diagonal} conductivity (per valley and spin)
\begin{widetext}
\begin{equation}
  \begin{array}[b]{r@{}l} \displaystyle
\sigma_g^{xy} = - \frac{\pi^2 e^3V_g \tau_-}{2h^2}
\lim_{\omega \rightarrow 0} \bigg[ \frac{1}
{1 + e^{-\beta(\mu/2 + \hbar \omega/4) }}
- \frac{1} {1 + e^{-\beta(\mu/2 - \hbar \omega/4) }} \bigg].
\end{array}
\end{equation}
\end{widetext}
The term $\sigma_g^{xy}$, like $\sigma^{xx}_{od}$ and like the term
analogous to $\sigma^{xx}_{od}$ in single-layer graphene,
\cite{UsGraphene} depends on the order of limits. Nevertheless, this
term is the \textit{only} contribution to the off-diagonal
conductivity, a fact that can clarify the correct order of limits.
Crystal symmetry and Onsager relations imply that an off-diagonal
conductivity $\sigma_{xy}$ requires time reversal symmetry breaking,
for example through the presence of a magnetic field. Therefore it
must be argued on physical grounds that an off-diagonal term in the
conductivity such as that found in the current work should not exist.
\cite{time} The only resolution to this is to take the limit
$\omega \rightarrow 0$ first. These observations imply that
$\sigma^{xx}_{od}$ in bilayer and single-layer graphene should be
zero.

We note that the findings of this section do not conflict with the
presence of a minimum conductivity $\sigma_{xx} = 4e^2/(\pi h)$ in
ballistic graphene, which has been measured experimentally.
\cite{Miao} The ballistic regime is qualitatively different in that
the mean free path greatly exceeds the sample size.

We thank E.~McCann, A.~H.~MacDonald, Y.~Barlas, S.~Adam, E.~Rossi
and E.~H.~Hwang for enlightening discussions. The research at
Argonne National Laboratory was supported by the US Department of
Energy, Office of Science, Office of Basic Energy Sciences, under
Contract No.\ DE-AC02-06CH11357.

\end{document}